\documentclass[aps,prb,twocolumn,groupedaddress,superscriptaddress,showpacs]{revtex4-1}
\usepackage{graphicx}
\usepackage{dcolumn}
\usepackage{bm}
\begin{document}


\title{Giant inelastic tunneling in epitaxial graphene mediated by localized states}


\author{J. \v{C}ervenka}
\affiliation{Institute of Physics, Academy of Sciences of the Czech Republic, v.v.i., Cukrovarnicka 10, 162 53 Prague, Czech Republic}
\affiliation{Physics Department, Eindhoven University of Technology, 5600 MB Eindhoven, The Netherlands}
\author{K. van de Ruit}
\author{C. F. J. Flipse}
\email[Corresponding author:~]{c.f.j.flipse@tue.nl}
\affiliation{Physics Department, Eindhoven University of Technology, 5600 MB Eindhoven, The Netherlands}


\date{\today}

\begin{abstract}
Local electronic structures of nanometer-sized patches of epitaxial graphene and its interface layer with SiC(0001) have been studied by atomically resolved scanning tunneling microscopy and spectroscopy. Localized states belonging to the interface layer of a graphene/SiC system show to have an essential influence on the electronic structure of graphene. Giant enhancement of inelastic tunneling, reaching 50\% of the total tunneling current, has been observed at the localized states on a nanometer-sized graphene monolayer surrounded by defects.
\end{abstract}


\maketitle


\section{\label{sec:level1}Introduction}

Epitaxially grown graphene on SiC offers one of the most promising platforms for applications in high speed electronic devices that might replace silicon in future.\cite{1} However, the quality of the two-dimensional electron gas in epitaxial graphene on SiC still falls behind the electronic quality of mechanically exfoliated graphene.\cite{1,2} While the maximum charge carrier mobility of epitaxial graphene on Si(0001) is in the order of 1000~cm$^2$/Vs at room temperature,\cite{2} the mobility reaches two orders of magnitude higher values in exfoliated graphene.\cite{3} Therefore a great deal of interest is focused on the understanding the differences between the electronic structure of exfoliated and epitaxial graphene and the consequences for potential applications. Since the crystallographic quality of both graphene layers seems to be equivalent,\cite{2} the interaction with the substrate remains the biggest unknown. This is mainly because of the complicated structure and electronic properties of the carbon rich graphene/SiC interface layer, which are still not fully understood.\cite{4,5,6}

In this paper, we present a local study of electronic and vibrational properties of nanometer-sized areas of a graphene monolayer grown on SiC(0001) and its $(6\sqrt{3}\times6\sqrt{3})R30^{\circ}$ interface layer by scanning tunneling microscopy (STM). Local scanning tunneling spectroscopy (STS) and inelastic electron tunneling spectroscopy (IETS) measurements have revealed unexpected phenomena in epitaxial graphene that could not be observed in spatially averaged measurements, which are usually reported in literature. Localized states of the interface layer protruding through the first graphene layer have caused giant enhancement of inelastic tunneling of electrons from graphene particularly on the places with localized electron states of the graphene/SiC interface layer. The inelastic phonon contribution for the out of plane graphene acoustic phonon at 70~mV has reached a gigantic 50\% of the total tunneling current. Our work reveals an unusual process of inelastic tunneling, which is principally different from previously reported phonon-mediated tunneling in mechanically cleaved graphene placed on SiO$_2$.\cite{7}

\section{Experimental}
The growth of atomically thin graphene samples was carried out in situ in ultra-high vacuum (UHV) on n-type 6H-SiC(0001) by thermal decomposition of Si at elevated temperatures. The growth process and have been done on a home-built electron-beam heater according to the preparation method described elsewhere.\cite{8} The sample temperature has been monitored by a pyrometer using emissivity 0.9. Owing to inhomogeneous heating of the sample by the e-beam heater, a mixture containing very small atomically flat areas (10-20~nm) of graphene mono-, bi- and interface layers has been produced as confirmed by low energy electron diffraction (LEED) and STM. Scanning tunneling microscopy experiments were performed in an Omicron GmbH LT-STM setup, working under UHV conditions (10$^{-11}$~mbar) at 5~K. Electrochemically etched W tips were used in the constant current STM mode. Scanning tunneling spectroscopy (STS) and inelastic electron tunneling spectroscopy (IETS) have been obtained by using two lock-in amplifiers and superimposing an alternating voltage reference signal with a frequency 990~Hz and amplitude 10~mV to the bias voltage applied to the sample.

\section{Results and discussions}

\subsection{Structural properties of grain boundaries}

Figure 1 shows spatially averaged STS curves on 0 interface, 1st and 2nd graphene layers on SiC, which are usually presented as local electronic structures of these layers.\cite{4} Even though the STS measurements have been obtained on areas with very small sizes (10-20~nm) that were surrounded by many large structural defects, they show comparable results to STS results reported by other groups on better quality graphene samples.\cite{4,9} However, averaging of STS curves is not appropriate in disordered systems such as the graphene/SiC(0001) system is, because it mixes incorrectly the local density of states (LDOS) at different locations. We illustrate this in Figure~\ref{fig2} by a series of atomically resolved STM images of a first graphene layer taken at different bias voltages. At low bias voltages ($\pm$50~mV), the characteristic graphene atomic structure together with the larger $(6\sqrt{3}\times6\sqrt{3})R30^{\circ}$ superstructure are visible, indicating a single graphene layer on SiC(0001).\cite{4,5,6,9} However, when the bias voltage is increased, bright dots start to appear until they fully dominate the STM pictures at higher voltages. Owing to these bias dependent topographic features, an average of STS spectra becomes bias dependent and therefore it does not reflect correctly an average of LDOS. 

\begin{figure} 
\includegraphics[width=3.3in]{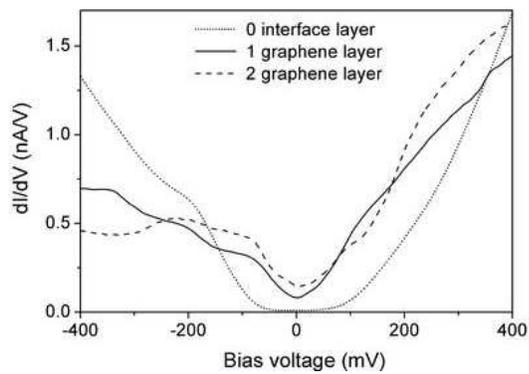}
   \caption{\label{fig1}Spatially averaged STS curves on the 0 interface, 1st and 2nd graphene layers on SiC(0001). Scanning parameters: $V=-200$~mV, $I=50$~pA for graphene monolayer and bilayer, and $V=-200$~mV, $I=5$~pA for the 0 layer.}
\end{figure}

The observed disordered bright features do not originate from the graphene layer but from the underlying interface layer as has been previously discussed by Rutter \textit{et al.}\cite{4} Graphene thus shows transparency at higher bias voltages to bright features from the lower interface layer in STM. Although the bright features in the zero interface layer seem to be disordered on local scale, they manifest the $(6\sqrt{3}\times6\sqrt{3})R30^{\circ}$  reconstruction with respect to the SiC crystal on larger scales as confirmed by large scale STM images and LEED.\cite{5,6} Interestingly, the positions of bright features are not the same in the filled and the empty states as symbolized by crosses and circles in Figure 2c,d. Circles and crosses point out the positions of the bright features in the filled states (-200~mV) and in the empty states (200~mV) respectively. Local STS measurements on top of these features on a graphene monolayer have revealed clear localized electron states at -200, -500 and 500~mV (see Figure~2e). On the other hand, STS spectra measured on regions with a graphene character (no bright features are observed in STM) have not shown any peaks in the LDOS. Similar localized states as on the first graphene layer have also been measured on bright features in the zero interface layer by STS in Figure~2f. The carbon rich interface layer has semiconducting properties with a 400~meV gap pinned in between the $\pm$200~mV localized states in accordance with previous STS measurements.\cite{4} The spatial extension of these localized states is in the order of 0.5~nm.

\begin{figure}[t] 
   \includegraphics[width=3.3in]{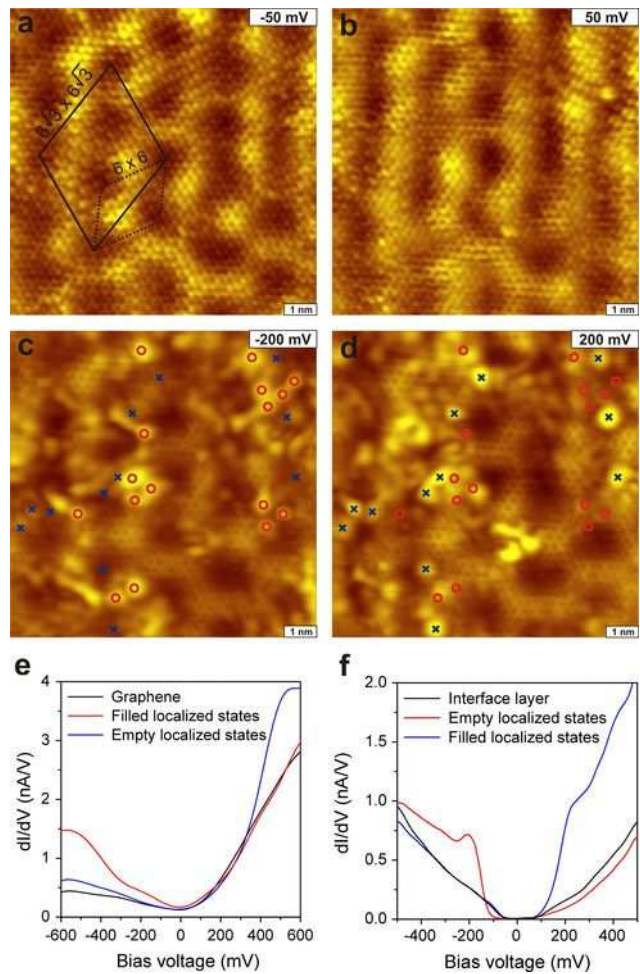}
   \caption{\label{fig2}(Color online) STM images of a $10\times10$~nm$^2$ area on single-layer graphene on SiC(0001) taken with $I=5$~pA and  $V=-50$~mV (a), 50~mV (b), -200~mV (c) and 200~mV (d). Circles point out the positions of the bright dots in the filled states and crosses in the empty states. (e) Three local characteristic $dI/dV$ spectra on monolayer graphene on SiC(0001) ($V=300$~mV and $I=41$~pA). (f) Three local characteristic $dI/dV$ spectra on the interface layer on SiC(0001) ($V=-200$~mV and $I=5$~pA). All STS curves have been averaged over 10 curves.}
\end{figure}

The origin of the localized states in the interface layer has been suggested to be either due to a different Si-C bonding in the interface layer consisting of covalently bonded graphene layer to the SiC(0001) surface\cite{Cervenka,10} or in the presence of Si adatoms.\cite{4} Both models propose correctly formation of localized states close to the Fermi energy. However, the first model is supported by angle resolved photoelectron spectroscopy (ARPES) studies on interface and graphene layers on SiC(0001)\cite{10,11} and by transferring of monolayer graphene to bilayer graphene after decoupling of the interface layer by H$_2$ intercalation.\cite{Riedl} In ARPES, the first graphene layer displayed well developed graphene $\pi$-bands extending up to the Fermi level, the interface layer exhibited semiconducting properties with absent $\pi$-bands.\cite{10} Two localized states at the binding energies 0.5~eV and 1.6~eV have been identified in the carbon rich interface layer with the $(6\sqrt{3}\times6\sqrt{3})R30^{\circ}$ reconstruction on the SiC(0001).\cite{10} Also the formation of empty electron states close to Fermi energy with a localized character has been observed in the graphitization study of SiC(0001) surface in inverse photoelectron spectroscopy.\cite{12} The localized states at $\pm$200~meV have not been identified in the photoemission experiments most probably because of their low intensities. Surprisingly, their energy coincides with a kink at 200~meV observed in the $\pi$-band dispersion near the $K$-point of monolayer graphene, whose origin has been suggested to be related to either electron-electron or electron-phonon interactions.\cite{13,14}

STS spectra of graphene monolayers and bilayers display an unexpected gap-like feature at the Fermi level (see Figure 1 or Ref. \cite{9}). From a thight-binding fit to photoemission data,\cite{6} however, one would not expect such a gap-like feature in STS because of the electron doping, which causes a shift of the Dirac point (the minimum in the graphene DOS) to -0.45~eV and -0.32~eV for monolayer and bilayer graphene layers respectively.\cite{14} Also transport experiments suggest a higher electron density on a monolayer graphene on SiC\cite{1,2} than on exfoliated graphene placed on SiO$_2$, where the Dirac point is in the vicinity of the Fermi energy.\cite{7,15} Recently the appearance of a gap-like feature at the Fermi-level on exfoliated graphene supported on a silicon oxide surface has been explained by the inability to tunnel into the  $\pi$-states due to a small tunneling probability at the Fermi-level.\cite{7} This has been overcome by the assistance of a phonon at 63~meV coupled with  $\sigma$-states, which made the tunneling possible at energies higher than the phonon energy.\cite{7} The experimental findings of Zhang \textit{et al.} have been supported by theoretical modeling of Wehling \textit{et al.}\cite{16} 

In Figure 3b, we show the observation of phonon contributions in IETS on a nanometer-sized monolayer graphene on SiC(0001). The inelastic tunneling features are observed as peaks (or dips) in the second derivative of current with respect to the voltage at the threshold where the electron energy associated with the bias voltage matches the oscillator energy. The $dI/dV$ and $d^2I/dV^2$ spectra in Figure 3 have been spatially averaged over 4096 points. Four inelastic peaks corresponding to out of plane acoustic graphene phonons at 16 and 70~mV can been identified in the $d^2I/dV^2$ spectrum on a graphene monolayer. Similar phonon modes at 16 and 58~mV have been found on graphite in IETS before.\cite{17} Phonon-induced inelastic tunneling in single molecules deposited on metal surfaces typically leads to conductivity changes in the order of only $\Delta\sigma/\sigma\approx1\%$,\cite{18} where the normalized change in differential conductance $\Delta\sigma/\sigma$ is obtained by normalizing the peak area in $d^2I/dV^2$ to conductance. The inelastic peak intensities in monolayer graphene on SiC are $\approx10\%$  for both phonon contributions at 16~mV and 70~mV (Figure~4). Surprisingly, the tunneling conductivity changed by a much larger factor 13 outside the gap-like feature on the exfoliated graphene.\cite{7} This has been explained by a different mechanism based on the phonon-mediated tunneling process which involves momentum-conserving virtual transitions between 2D electron bands in graphene.

\begin{figure}[t] 
   \includegraphics[width=3.3in]{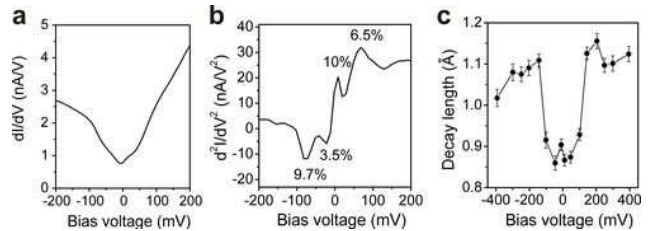}
   \caption{\label{fig3}Averaged $dI/dV$ (a) and $d^2I/dV^2$ (b) spectra taken on graphene monolayer on SiC(0001) and its tunnel current decay length $\lambda$ as a function of bias voltage (c). An inelastic peak intensity $\Delta\sigma/\sigma$ is indicated at the each attributed phonon peak in $d^2I/dV^2$. Scanning parameters: $64\times64$ grid, $V=50$~mV and $I=70$~pA. The decay lengths have been determined from $I(z)$ spectroscopy at fixed bias voltage by fitting it by an exponential function $I(z)=exp(-z/\lambda)$. Error bars represent standard deviations of the measurements.}
\end{figure}

The mechanism of the phonon-assisted tunneling in exfoliated graphene was supported by observation of bias dependent wavefunction spatial decay rates, where the tunnel decay length inside and outside the gap has been observed to be 0.25~$\AA$ and 0.45~$\AA$ respectively.\cite{7} Bias dependent wavefunction spatial decay rates in monolayer graphene grown on SiC are depicted in Figure~4c. The decay length $\lambda$  has been determined from $I(z)$ spectroscopy performed at fixed bias voltage $V$ by fitting it to an exponential function $I(z)=exp(-z/\lambda)$. Similarly like on exfoliated graphene, two different decay rates have been observed inside and outside the gap-like feature bounded in the $\pm$100~mV region, $\lambda_{IN}=0.89~\AA$  and $\lambda_{OUT}=1.1~\AA$. 

Although the results measured on epitaxial graphene in Figure 3 look similar to the data by Zhang \textit{et al.} measured on exfoliated graphene,\cite{7} the mechanism is different. Firstly, both out of plane acoustic phonon contributions at 16 mV and 70 mV have similar intensities but only the latter phonon can assist the virtual tunneling to $\sigma$ electrons since it has the right momentum because it is centered at the  $K/K'$ points, whereas the other out-of-plane acoustic phonon at 16~mV cannot play the same role because it is located at the $\Gamma$ point. Secondly, the tunneling decay rates are observed to change exactly at the edge of the gap of the interface layer (see Figure 3), whose states are known to have a large spatial extension since they are seen in STM even upon formation of two graphene layers above the SiC interface. Finally, the most important fact that disproves the phonon assisted tunneling in epitaxial graphene on SiC is a spatially inhomogeneous character of the inelastic contribution. 

\begin{figure}[t] 
   \includegraphics[width=3.3in]{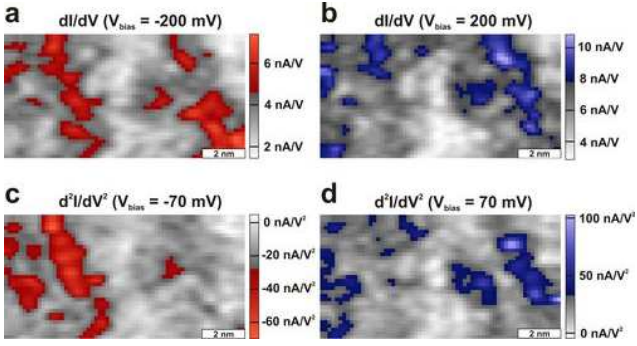}
   \caption{\label{fig4}(Color online) $dI/dV$ (a,b) and $d^2I/dV^2$ (c,d) maps at constant bias voltage indicated in the right top corner of a $12\times8$ nm$^2$ area on graphene monolayer on SiC(0001). Red regions indicate high intensity inelastic phonon contributions at -70~mV and high $dI/dV$ at -200~mV and blue regions mark out high intensity $d^2I/dV^2$ at 70~mV and $dI/dV$ at 200~mV. Scanning parameters: $V=50$~mV and $I=50$~pA.}
\end{figure}

To illustrate the spatial dependence of inelastic tunneling intensity, we show simultaneously measured $dI/dV$ and $d^2I/dV^2$ maps on a graphene monolayer on SiC(0001) in Figure~4. The $d^2I/dV^2$  images depict intensities of the inelastic peak contribution of the phonon mode at $\pm$70~mV and the $dI/dV$  maps portray intensities of the localized states at $\pm$200~mV. The places of the high inelastic peak intensity coincide with the places where the $\pm$200~mV localized states are observed in the $dI/dV$ maps. For this reason, high intensity regions in Figure 4 have been highlighted by red and blue color in negative and positive bias voltages respectively to highlight the correlation between $dI/dV(\pm$200~mV$)$ and $d^2I/dV^2(\pm$70~mV$)$ maps. 

The IETS peak intensities vary spatially by a large factor in $d^2I/dV^2$ maps, up to 50 among some places, as seen by the difference between the values of red/blue and gray regions. The regions with high IETS intensities are found at different locations in positive and negative bias voltage, similar to the bright features in Figure~1. This inhomogeneous asymmetry can be also seen on three characteristic local  $dI/dV$ and $d^2I/dV^2$ spectra depicted in Figure~5. These spectra have been averaged only over 10 local measurements, therefore they exhibit a larger noise level in comparison to the spatially averaged IETS spectra. An IETS curve measured on a position with a high IETS intensity at -70~mV (Figure 5a) shows a gigantic inelastic feature reaching $\Delta\sigma/\sigma\approx50\%$ in negative bias voltage, while the IETS peak in positive voltage is half of this size. Such high IETS signals have been observed predominantly at positions with high $dI/dV$ intensities at -200~mV. These places most probably correspond to the localized states at -200~mV on the first and zero graphene layers. Moreover, a second harmonic phonon mode at -140~mV is observed in $d^2I/dV^2$ with an intensity approximately 5 times smaller than the intensity of the first harmonic mode. Similar results have been observed on places with a high inelastic peak at +70~mV that are located at position with a high $dI/dV$ at 200~mV, implying a connection with localized states of the graphene monolayer in the empty states. In this case, an enormous first order inelastic peak together with the second harmonic contribution has been observed in the positive bias voltage. On the other hand, IETS spectra obtained on locations free of localized states (Figure 5c) have demonstrated relatively low intensity phonon contributions (10\%) for both 16 and 70~mV out-of plane phonons. No second order phonon modes could be seen in these IETS spectra. Important is to note that one should be careful in relating the high intensity $dI/dV$ regions at $\pm$200~mV with localized states since an increase in $dI/dV$ can also be caused by high intensity IETS peaks at  $\pm$70~mV. However, since the presence of localized states have also been independently proved by other STM groups on a monolayer graphene,\cite{9} it is highly probable that high intensity $dI/dV$ correlates with localized states at  $\pm$200~mV originating in the graphene/SiC(0001) interface layer.

\begin{figure}[t] 
   \includegraphics[width=3.3in]{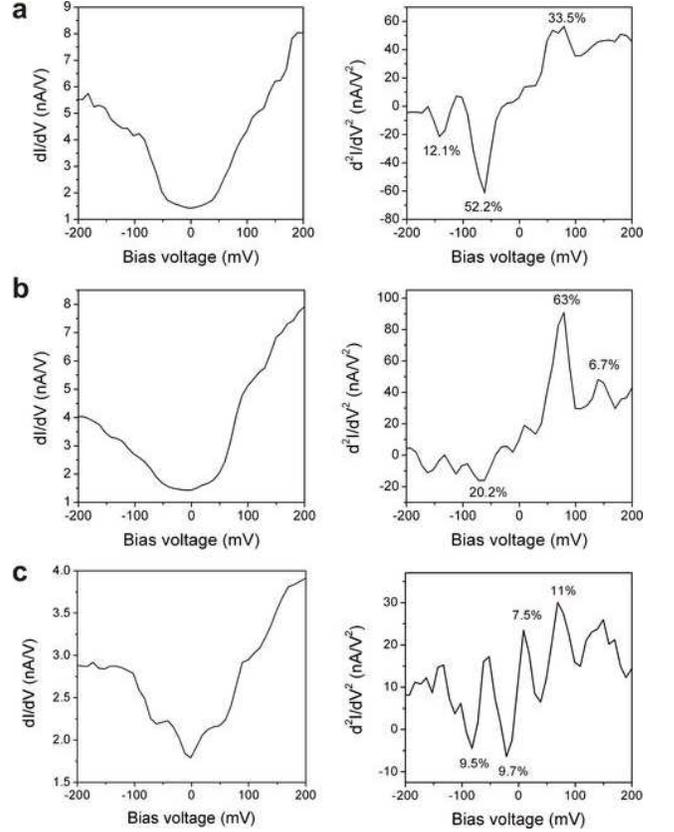}
   \caption{\label{fig5} $dI/dV$ and $d^2I/dV^2$ spectra on graphene monolayer on SiC taken at $V=50$~mV and $I=50$~pA. The spectra represent typical individual $dI/dV$ and $d^2I/dV^2$ curves obtained in red reagions (a), blue regions (b) and gray regions (c) in Figure 4. Inelastic peak intensities $\Delta\sigma/\sigma$ are indicated at the each attributed phonon peak in $d^2I/dV^2$.}
\end{figure}

Typical inelastic tunneling experiments give rise to phonon fingerprints that reach only a few percent of the total tunneling current, however, we observe unexpectedly giant signals as high as 50\% in the presence of localized states. This could indicate a strong electron-phonon (e-ph) coupling strength ($\lambda$), but calculations by Park \textit{et al.}\cite{19} have shown values for $\lambda$ in the order of 0.05 at 200~meV binding energy. This relative small value cannot explain the large IETS intensity. The fact that the large IETS intensities coincide with the positions of sharp localized electron states at $\pm$200~mV is supported by the DFT calculation results of Atta-Fynn \textit{et al.}\cite{20} In this study, localized electron states stemming from defects or topological disorder exhibited an anomalously large e-ph coupling.\cite{20} Hence, the observed localized states probably enhance the e-ph coupling, resulting in a larger IETS intensity. However, the presence of localized states might not be the only criterion of giant IETS contributions because the IETS have been measured on a very small graphene regions (10-20~nm) confined among many structural defects. Therefore, there seem to be two conditions for the giant enhancement of the IETS data: both the influence of the localized states at $\pm$200 mV and the presence of structural defects. The structural defects are known to play a very important role in the scattering of electrons, which is an additional contribution for localization, thus causing together with localized states an anomalously large e-ph coupling.
 
In addition, in the $dI/dV$ spectra, higher harmonics are observed equidistantly spaced with the value of the vibration. Higher harmonics so called phonon (vibrational) side bands have been observed occasionally in scanning tunneling experiments in the resonant tunneling regime.\cite{21} The conditions for resonant tunneling are discussed in detail by Galperin \textit{et al.}:\cite{22} the higher order vibronic levels become visible if the tunneling electron stays relatively long on the molecule compared to the dephasing time and the localized electron state coincides with a vibration level. This happens if the chemical interaction between electron state and molecule is relatively small, resulting in a narrow broadening of the vibronic level.\cite{22} As is shown, at $\pm$200~mV, localized electron states can couple with the out-of-plane phonons from graphene. Because the $dI/dV$ data do not show periodic phonon peaks equidistantly around $\pm$200~mV, the resonant tunneling channel is not related to the localized states at $\pm$200~mV. Resonant tunneling through localized states at the Fermi-level is difficult to determine because of the pseudogap, but it would be highly probable if the origin of the pseudogap is of many-body character, characterized by electron-electron and electron-phonon interactions.\cite{23}

\section{Conclusions}

In conclusion, a giant inelastic tunneling process has been observed in epitaxial graphene on SiC(0001) in scanning tunneling experiments. The inelastic tunneling channel reached half of the total tunneling current. The mechanism of the giant tunneling is connected with the presence of sharp localized states originating in the interface with SiC and strong electron-phonon coupling in graphene near a structural defect. 

\begin{acknowledgments}
The authors are grateful to Thomas Seyller for providing SiC samples and for fruitful discussion. This research was financially supported by Nanoned.
\end{acknowledgments}


\end{document}